\documentclass[twocolumn,preprintnumbers,amsmath,amssymb,amsfonts,prb,floatfix]{revtex4}

\usepackage{graphicx} 
\usepackage{dcolumn} 
\usepackage{bm} 

\begin{document}

\title{Effect of quantum confinement of surface electrons on an
atomic motion on nanoislands}
\author{A.S.~Smirnov$^{(1,2)}$}
\author{N.N.~Negulyaev$^{(1)}$}
\author{L.~Niebergall$^{(3)}$}
\author{W.~Hergert$^{(1)}$}
\author{A.M.~Saletsky$^{(2)}$}
\author{V.S.~Stepanyuk$^{(3)}$}
\address{$^{(1)}$ Fachbereich Physik, Martin-Luther-Universit\"at,
Halle-Wittenberg, Friedemann-Bach-Platz 6, D-06099 Halle, Germany}
\address{$^{(2)}$ Faculty of Physics, Moscow State University,
119899 Moscow, Russia}
\address{$^{(3)}$ Max-Planck-Institut f\"ur Mikrostrukturphysik,
Weinberg 2, D-06120 Halle, Germany}

\begin{abstract}
\emph{Ab initio} calculations and kinetic Monte Carlo simulations
demonstrate that the quantum confinement of surface electrons to
nanoislands can significantly affect the growth process at low
temperatures. Formation of empty zones and orbits of an adatom
motion is demonstrated for Cu nanoislands on Cu(111).
\end{abstract}

\maketitle

Noble (111) metal surfaces support Shockley surface
states\cite{c1}. These electronic states arise in the inverted $L$
gap of a metal band structure. Electrons occupying the
Shockley-type surface states form a 2D nearly free electron gas
confined in the vicinity of the top layer\cite{c2}. Surface
electrons scatter at point defects\cite{c3}, adsorbates\cite{c4},
steps\cite{c5}, leading to standing waves of local density of
states (LDOS), which can be detected using scanning tunneling
microscope (STM) technique. These standing-wave patterns contain
important information about properties of scattering sites and
interaction between the surface electrons and scattering centers.
Particularly, scattering of the surface states at adatoms leads to
the standing-wave patterns around the adsorbates and to an
indirect long-range interaction (LRI) between adatoms\cite{c6}.
While the first experimental evidence of such type of interactions
has been reported 35 years ago\cite{c7}, quantitatively they have
been resolved only recently by means of low-temperature STM
studies\cite{c8_0,c8}. These remarkable observations have opened a
door to creation of new types of macroscopic-ordered 2D\cite{c9}
and 1D\cite{c10} nanostructures stabilized by the surface
electrons.

Several fascinating phenomena occur if surface electrons are
confined to closed nanostructures, like corrals, vacancy holes or
clusters. Experiments of Manoharan et al.,\cite{c11} and later
\textit{ab initio} calculations\cite{c12} have demonstrated that
the quantum confinement inside corrals induces a mirage effect. It
is also possible to tailor spin-polarization of surface-state
electrons and exchange interaction between magnetic adatoms within
the confined nanostructures\cite{c13}. Confined surface electrons
inside quantum resonators alter diffusion at low temperatures,
leading to atomic self-organization\cite{c14}. To the best of our
knowledge Li et al.,\cite{c15} performed the first quantitative
investigation of the quantum confinement of surface electrons on
nanoscale Ag islands on Ag(111) by means of STM. These studies
revealed the validity of the confinement picture down to the
smallest of island sizes. It was proposed\cite{c15} that the
quantum confinement can arise not only on Ag(111), but on Cu(111)
and Au(111). Observation of quantized electronic states in vacancy
islands on Cu(111) has been recently reported\cite{c16}.

In this paper we demonstrate the effect of confined surface
electrons on atomic motion in close proximity and on top of
nanoislands. The quantum confinement is studied by means of the
first principles Korringa-Kohn-Rostocker (KKR) Green's function
method. Our kinetic Monte Carlo (kMC) simulations reveal that
confinement-induced electronic states around and on top of
nanoislands significantly affect atomic diffusion. Formation of
empty zones and orbits of adatom motion are shown. We demonstrate
that the quantum confinement dramatically effects the growth
process of nanoislands at low temperatures.

Within our study calculations of the electronic interaction
between an adatom and a nanoisland are performed using density
functional theory in local spin density approximation by means of
the KKR Green's function method\cite{c17}. The basic idea of this
method is a hierarchical scheme for the construction of the
Green's function of nanostructures on a surface by means of
successive applications of Dyson's equation. The bulk, surface and
impurity problems are consequently treated with a perturbative
approach. At each stage a fully self-consistent Green's function
is obtained, which is then used as a reference for the next step.
We treat a surface as a 2D perturbation of an ideal bulk with a
slab of vacuum. Taking into account the translational symmetry of
the surface geometry, the Green's functions are formulated in
momentum space. Adatom and nanoisland are considered as a
perturbation of a clean surface. These calculations are performed
in real space. Previous studies have demonstrated that this method
allows one to obtain with a good accuracy a spatial modulation of
LDOS and interaction energy in closed
nanostructures\cite{c12,c13,c16}.

As a model system we consider a hexagonal Cu nanoisland placed in
fcc hollow sites on a Cu(111) surface\cite{c18} and study
diffusion of a Cu adatom near and on top of it. Each edge of the
island has length of 11 atoms (2.6 nm). Two different kinds of
close-packed steps can be distinguished in the nanoisland: the
$\{100\}$-microfaceted ($A$) step and the $\{111\}$-microfaceted
($B$) step. We neglect the elastic interaction between the adatom
and the island\cite{c19}.

\begin{figure}
\begin{center}
\includegraphics[width=8.5cm]{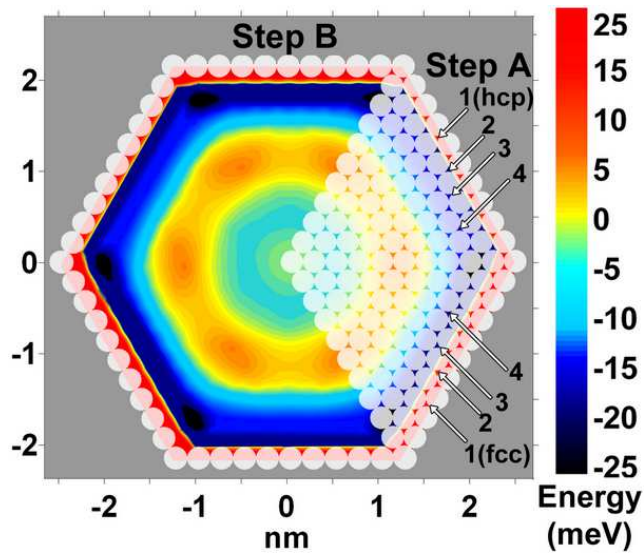}
\caption{(Color online) The 2D map of the interaction potential
between a Cu nanoisland and a Cu adatom placed on top of it.
Semi-transparent white circles demonstrate the edges of the island
and 1$/$3 part of it. The rows of hollow sites are numerated from
the steps (numbers 1-4).}\label{fig1}
\end{center}
\end{figure}

At first, we concentrate on the quantum confinement on top of the
island. Fig.1 presents the 2D potential energy map of a Cu adatom
placed in different hollow sites. The interaction potential
exhibits an oscillatory behavior if one recedes from an edge of
the nanoisland towards its center along the perpendicular bisector
to the edge. When the Cu adatom is located in the first row of
hollow sites near an $A$ or a $B$ step (Fig.1), the interaction is
repulsive being larger than 30 meV. We attribute this effect due
to redistribution of the electron-charge density at steps as was
suggested a long time ago by Smoluchowski\cite{c23}. When the Cu
adatom is located in the second row of hollow sites near $B$ step
the interaction is also repulsive being 10 meV. Our calculations
reveal the attractive minima of -19 meV at 1.5 $\AA$ from an $A$
step (the second row of hollow sites) and -20 meV at 2.9 $\AA$
from a $B$ step (the third row). Difference in the positions of
the minima with respect to the island edges appears due to the
different types of atomic packing at $A$ and $B$ steps. The
existence of attractive potential in a vicinity of a descending Cu
step is confirmed by experimental observations of Repp et
al.,\cite{c8_0}. One of the most fascinating effects of confined
surface electrons is quantum interference at the corners (black
spots in Fig.1). The attractive potential in these spots is deeper
than at steps being -25 meV. Similar phenomenon has been already
observed for the scattering of surface-state electrons in a
'hand-made' triangular corral constructed of Ag adatoms on a
Ag(111) surface\cite{c25}.

\begin{figure}
\begin{center}
\includegraphics[width=8.5cm]{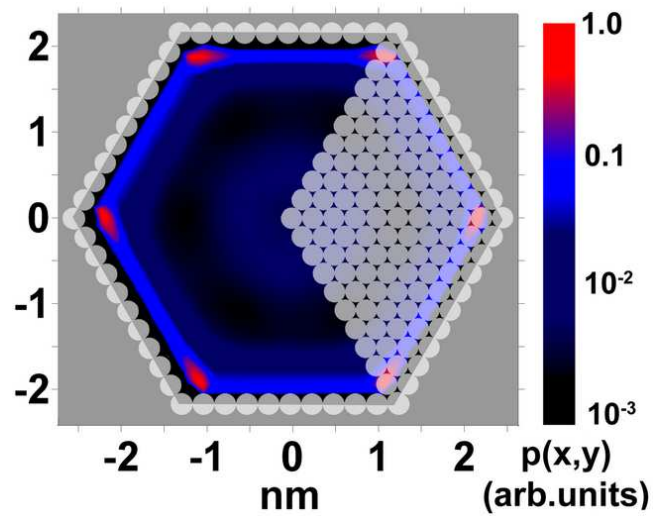}
\caption{(Color online) Atomic diffusion on top of the nanoisland:
results of the kMC simulations for a single Cu adatom. The
probability to find a randomly deposited Cu atom is calculated at
60 K. Semi-transparent white circles demonstrate the edges of the
island and 1$/$3 part of its surface. The logarithmic scale is
used.}\label{fig2}
\end{center}
\end{figure}

In order to give a clear evidence that atomic motion on top of the
nanoscale island is substantially modified compared to that on an
open surface, we perform the kMC study\cite{c26} of behavior of a
Cu adatom randomly deposited on top of the island. The hop rate of
an adatom from site $k$ to site $j$ is calculated using the
formulae: $\nu=\nu_0exp(-E_{k \to j}/k_BT)$, where $T$ is the
substrate temperature, $\nu_0$ is the attempt frequency (10$^{12}$
sec$^{-1}$ (see Ref.\cite{c27})) and $k_B$ is the Boltzmann
factor. The influence of the electronic interaction
'adatom-island' (Fig.1) is included in the diffusion barrier:
$E_{k \to j}=E_D+0.5(E_j-E_k)$, where $E_D$ is the energy barrier
for a Cu atom on a clean surface (40 meV\cite{c20}) and $E_{k(j)}$
is the magnitude of interaction, when the adatom is in the hollow
site $k(j)$. Previous studies\cite{c27} have confirmed the
applicability of this approach.

The adatom probability distribution on top of the island is shown
in Fig.2. Contrary to the traditional view, the adatom does not
diffuse as it does on a flat surface. Regions marked with the red
color correspond to the energy minima of -25 meV in the corners of
the island (Fig.1). The probability to find the adatom in these
six spots is significantly higher than that one along the island
edges. At the same time a localization of the adatom in a narrow
stripe along the steps is more preferable than in the center of
the island. Formation of empty zones and orbits of adatom motion
takes place.

\begin{figure}
\begin{center}
\includegraphics[width=8.9cm]{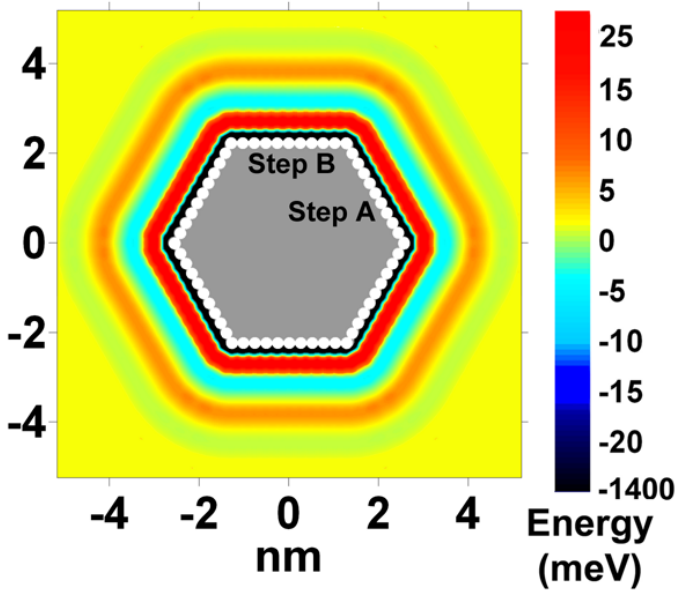}
\caption{(Color online) 2D interaction potential between a Cu
nanoisland and a Cu adatom placed in its vicinity. The edge of the
island is marked with the white circles.}\label{fig3}
\end{center}
\end{figure}

\begin{figure}
\begin{center}
\includegraphics[width=6.2cm]{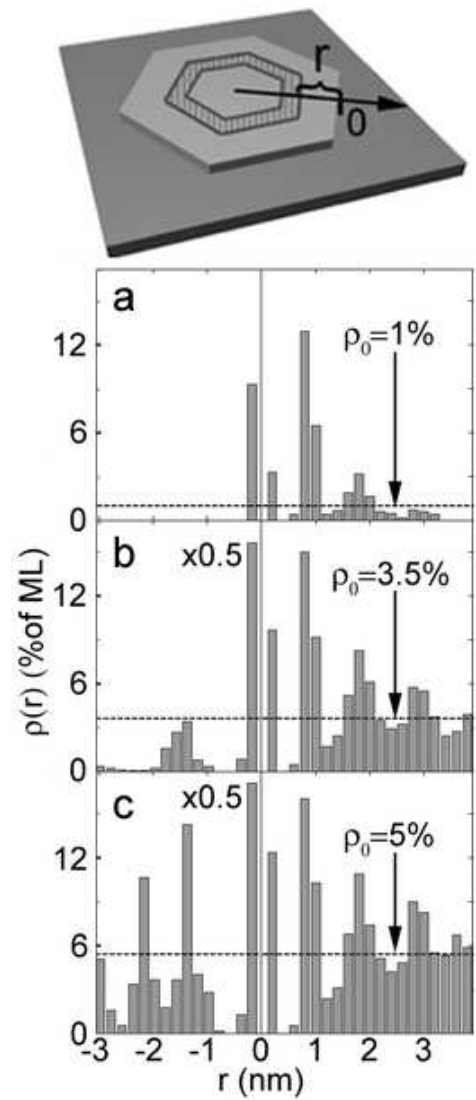}
\caption{Surface density of adatoms $\rho(r)$ on top of the
nanoisland and around it as a function of the distance $r$ to the
nearest island edge. Function $\rho(r)$ is a number of adatoms
over the square of a hexagonal concentric ring (marked with a
shadow). Positive values of $r$ correspond to the region outside
the island, while negative ones to that on top of it. The results
of the kMC simulations of self-assembly of Cu adatoms at a
presence of Cu nanoisland on Cu(111) are demonstrated. The Cu
adatoms are deposited simultaneously at 14 K; the coverage
$\rho_0$ is: a) 1$\%$ of ML, b) 3.5$\%$ of ML, c) 5$\%$ of ML. In
Figs. b),c) peaks at $r$ = -2 $\AA$ are rescaled by factor $0.5$.
Dashed lines correspond to the uniform atomic density at a given
coverage.}\label{fig4}
\end{center}
\end{figure}

Now we turn to the effect of the quantum confinement around the
island. The 2D potential energy map caused by the
substrate-mediated electronic interaction between the adatom and
the nanoisland is presented in Fig.3. The interaction has an
oscillatory behavior, decaying to zero if one recedes from the
island along the perpendicular bisector to a step edge. The
interaction is strongly attractive (about -1.4 eV), when the Cu
adatom is at 2.2 $\AA$ from the step; it corresponds to
aggregation of the adatom and the island. Our calculations reveal
a concentric repulsive ring of 25 meV at 4-7 $\AA$ from the edge
of the island. This barrier prevents atomic diffusion towards the
steps at low temperatures. There is a ring of attractive
interaction of -5 meV at 8-14 $\AA$, and a ring of repulsive
interaction of 2 meV at 15-20 $\AA$. The next minima and maxima of
energy are less than 1 meV. Concentric rings of negative and
positive energies (Fig.3) lead to the formation of empty zones and
'allowed' orbits. Particularly, at low temperatures a migrating
adatom prefers to be at about 9 $\AA$ from the steps, while the
probability to find an adatom at 4-7 $\AA$ from the edge of the
nanoisland is very low.

The quantum confinement has a dramatic effect on growth of small
clusters at low temperatures. To illustrate this statement we
perform the kMC simulations of self-organization of Cu atoms on
Cu(111) at 14 K for different coverages of deposited atoms
$\rho_0$. Within our study the LRI between Cu adatoms\cite{c29} is
described by a pairwise summation. Previous studies have confirmed
that this approximation is well-justified at large interatomic
separations\cite{c27}.

In Fig.4 we present histograms of statistically averaged density
of adatoms on top of the island and around it as function of the
distance between the adatom and the nearest step. Positive
distances $r$ correspond to the area outside the island, while
negative ones to that on top of it. At coverage $\rho_0$ = 1$\%$
of ML (Fig.4a) there is only one peak of the density on top of the
island. It indicates that all adatoms are localized at separations
of 1.5-3 $\AA$ from the steps, forming the first inner orbit. The
first peak outside the island (at $r$ = 2 $\AA$) corresponds to
atoms nucleated to the island during deposition. The second
maximum of atomic density outside the island is located at $r$ = 9
$\AA$: these atoms form the first outer orbit. The probability to
find an adatom on the second orbit (at $r$ = 17-22 $\AA$) is much
smaller. The electronic interaction 'adatom-nanoisland' at such
distances is less than 2 meV (Fig.3), and the second orbit is
stabilized by the LRI with the adatoms placed on the first orbit.

Our simulations indicate that effect of the quantum confinement on
atomic self-organization remains significant in a wide range of
coverages. The results for $\rho_0$ = 3.5$\%$ of ML are presented
in Fig.4b and for $\rho_0$ = 5$\%$ of ML are given in Fig.4c. With
increasing the coverage the probability of the occupation of the
first orbit on top of the island (at $r$$\sim$-2 $\AA$) increases.
The second and the third inner orbits (at $r$$\sim$-12 $\AA$ and
at $r$$\sim$-22 $\AA$ respectively) are formed. The similar
phenomenon occurs around the island: adatoms form orbits at
$r$$\sim$18 $\AA$ and $r$$\sim$28 $\AA$, while the atomic density
between the orbits is low. Two Cu atoms located on the same orbit
are separated by 11-12 $\AA$, corresponding to the position of the
local minimum of the LRI between two Cu adatoms\cite{c29}.

Finally, we would like to note that empty zones were found in
several experimental works of Ehrlich and co-workers. In
Ref.\cite{c30,c31} diffusion of an Ir adatom around an Ir cluster
on Ir(111) was studied. Empty zone was resolved at separations of
3-7 $\AA$ from the edge of the cluster, while the preferable
distance between the migrating atom and the step edges was 9-12
$\AA$. With increasing the temperature the adatom overcame the
empty zone and nucleated with the cluster. The anisotropy
diffusion was also observed for an Ir adatom placed on top of an
Ir cluster on Ir(111). The Ir atom was localized near the edge of
the cluster with a higher probability than in its central
region\cite{c31,c32}. Diffusion of a Pt atom on top of a Pt
cluster on Pt(111) was a subject of study in Ref.\cite{c33}. It
was found that empty zone (of width of 7 $\AA$) divides the region
with uniform atomic distribution in the center of the cluster from
the orbit near steps. During migration along the step edges, the
preferable location of a Pt adatom was in the corners of the
cluster. Results of our study for Cu suggest that phenomena
observed by Ehrlich and co-workers could be promoted by the
quantum interference of electrons scattered on the cluster, on one
hand, and on the adatom, on another hand. However, in order to
give a precise answer to this question, further theoretical
investigations are required.

In conclusion, we have performed \emph{ab initio} study of the
quantum confinement of surface electrons on top of nanoscale Cu
islands on Cu(111) and around them. Formation of concentric rings
with positive and negative interaction energies is revealed
outside and on top of a nanoisland. This phenomenon substantially
modify the atomic diffusion at low temperatures, leading to empty
zones and orbits of adatom motion. Our results indicate on a
profound role of the quantum confinement of surface electrons
during growth of metal nanoclusters at low temperatures.

This work was supported by Deutsche Forschungsgemeinschaft
(SPP1153, SPP1165).


\begin{thebibliography}{00}
\bibitem{c19} Using the molecular static method with many-body
interatomic potentials for a system of Cu atoms\cite{c20} we found
that elastic interaction between the adatom and the nanoisland at
any separations is negligible small compared to the electronic
interaction.
\end{thebibliography}
\end{document}